\documentclass[10pt,a4paper]{raa}           
\usepackage{graphicx,times}
\usepackage{natbib}
\usepackage{amssymb,amsmath}
\bibpunct{(}{)}{;}{a}{}{,}

\begin{document}

 \title{Statistical analysis on Dynamic Fibrils observed from NST/BBSO observations 
$^*$
}

 \volnopage{ {\bf 2012} Vol.\ {\bf X} No. {\bf XX}, 000--000}
   \setcounter{page}{1}

     \author{T G Priya
        \inst{1,2}
     \and Su Jiangtao
        \inst{1,2}
     \and Jie Chen
             \inst{1,3}   
     \and Deng Yuanyong
        \inst{1}
      \and Debi Prasad Choudhury
         \inst{4}
         }
     

   \institute{  Key Laboratory of Solar Activity, National Astronomical
   Observatories, Chinese Academy of Sciences, Beijing 100012, China; {\it priyatg@nao.cas.cn}\\
        \and
             University of Chinese Academy of Sciences 19 A Yuquan Rd, Shijingshan District, Beijing 100049, China \\
\and
State Key Laboratory for Space Weather, Center for Space Science and Applied Research, Chinese Academy of Sciences, Beijing 100190, China\\
\and 
California State University, Northridge, CA 91330, USA \\
}

\abstract
{We present the results obtained from the analysis of dynamic fibrils in NOAA active region (AR) 12132, using high resolution H$\alpha$ observations from New Solar Telescope operating at BBSO. The dynamic fibrils are seen to be moving up and down, and most of these dynamic fibrils are periodic and have jet like appearance. 
 We found from our observations that the fibrils follows perfect parabolic paths at the most in many cases. A statistical measure on the properties of the parabolic paths showing an analysis on deceleration, maximum velocity, duration and kinetic energy of these fibrils is presented here.  We found the average  maximum velocity to be around 15 km s$^{-1}$ and mean deceleration to be around 100 m s$^{-2}$. The deceleration observed appears to be a fraction of gravity of sun and is not compatible with the path of ballistic at the gravity of sun. We found positive correlation between  the deceleration and the maximum velocity. This correlation is consistent with the simulations done earlier on magnetoacoustic shock waves propagating upward. 
} 

   \authorrunning{Priya et al. }            
   \titlerunning{Dynamic fibrils}  
   \maketitle

%
\section{Introduction}           
\label{sect:intro}

The chromosphere that lies between the photosphere and the hot corona is  often reveals jet like features, which are small in size and has very short lifetimes.  Though there have been a lot of
confusions with the relationship between the quiet sun limb
spicules, mottles observed on the quiet sundisk and the
dynamic fibrils (DFs)\citep{1992A&A...264..236G},  the three phenomena are related \citep{1994A&A...290..285T}.  The physical process of chromosphere is not well understood, especially the mass and energy transportation in dynamic fibrils. \citep[see][] {2012RSPTA.370.3129R}.The flows (up and down) from photosphere to chromosphere and the higher layer corona flaunt a variety of  oscillation modes like three- minute, five-minute and the seven-minute oscillations.  \citet{2014ApJ...786..137T}  presented strong evidence to the shock behavior of sunspot oscillations for the first time in sunspots in the transition region (TR) and in the chromosphere based on the IRIS observations.  The 5-minute oscillations \citep{1962ApJ...135..474L} happening over the sunspots have been studied extensively over a period of time and have been found that there is a decrease in amplitude with height and not easily detected in the higher layers above photosphere. The 3-minute oscillations could be due to the mode conversion of  magnetoacoustic waves \citep{2012A&A...539L...4S}.  The fibrilar structure of chromosphere reveals that magnetic fields play a significant role \citep{1908ApJ....28..315H}. The jet like features which are also known as spicules, mottles, and the DFs are observed at limb, quiet sun disk and in AR respectively. \\ 
Our main concentration is on the observations of DFs. The dynamics of DFs and its imprint on TR, and its oscillations are extensively studied \citep{1999SoPh..190..419D,2003ApJ...595L..63D,2003ApJ...590..502D,2004Natur.430..536D}.  A remarkable power of oscillations in the Dfs were determined by  \citet{2003ApJ...595L..63D}. This was discovered using high resolution data from SST. \citet{2004Natur.430..536D} and \citet{2006RSPTA.364..383D} build up a model which explains the reason for DFs to be the inclination of magnetic field lines. This inclination reduce acoustic cut off frequency permitting the longer period so called,  p-mode waves, to leak into the atmosphere where it forms shock and these shocks are the reason for driving the DFs. This was further investigated and proved by \citet{2006ApJ...647L..73H} using the similar data from SST that the DFs are caused by magnetoacoustic shock waves. 
In this paper we performed an analysis on the motion of DFs using New Solar Telescope \citep[NST,][]{2010AN....331..636C} (now known as GST, Goode Solar Telecope, after July 2017 ) operating at Big Bear Solar Observatory (BBSO) by measuring its properties and described its temporal evolution. Our results supports the fact that the DFs are formed due to the shock waves in chromosphere that are driven by the flows that were convective and photospheric oscillations as reported by De Pontieu et al. (2004). Through analysis of dataset taken at NST, we find that the DFs and its properties could be a result of shock waves that are generated from the p-mode propagating upward.

   \begin{figure}
           \centering
           \includegraphics[width=15.0cm, angle=0]{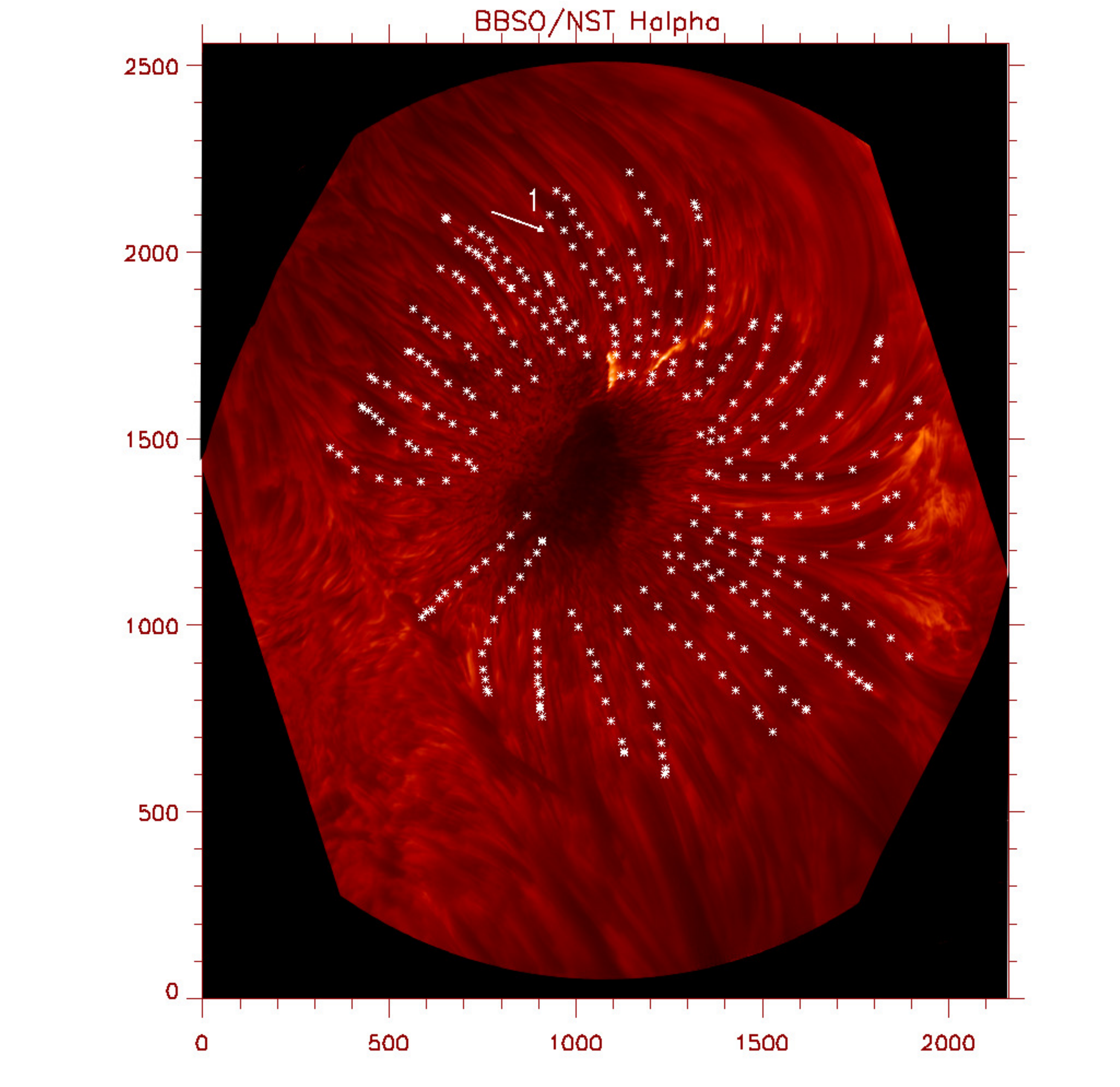}
           \caption{\normalsize H$\alpha$ line core image showing the paths of DFs overplotted. The DF marked as '1' is used for further analysis. Tick marks are in Pixels. }
           \label{Fig1}
        \end{figure}
 
\section{Observations and data processing}
\label{sect:Obs}

The sunspot of NOAA AR 12132, located at S09E08 on 2014 August 5 from BBSO  \citep[NST,][]{2010AN....331..636C}, was chosen for studying the DFs (see Figure 1). The pointer  was (270$\arcsec$, −395$\arcsec$), targeting the sunspot of AR 12132. The observations were performed using NST during 18:20 UT$-$19:20 UT. The Chromospheric images were acquired every 23 s by scanning of the H$\alpha$ spectral line from the blue wing -1 \AA{} to red wing +1 \AA{} with a step of size 0.2 \AA{}. The FOV is about 70$\arcsec$, and has a pixel scale of 0.029$\arcsec$ pixel$^{-1}$. The data is used to investigate the umbral oscillations in chromospheric sunspot at  different solar altitudes observed on August 5. A  combination of a  5 \AA{} interference filter and a Fabry–Pérot etalon is used in the  VIS to get a  bandpass of  0.07 \AA{} at the H$\alpha$ line. We chose the first H$\alpha$ -1.0 \AA{} image as a  reference image to align all other images in this passband. The relative  shifts were recorded, and used  to register the images in the other passbands of H$\alpha$.

     \section{Data analysis and results}\label{sec:analysis} 
            \begin{figure*}
            \centering
            \includegraphics[width=14.0cm, angle=0]{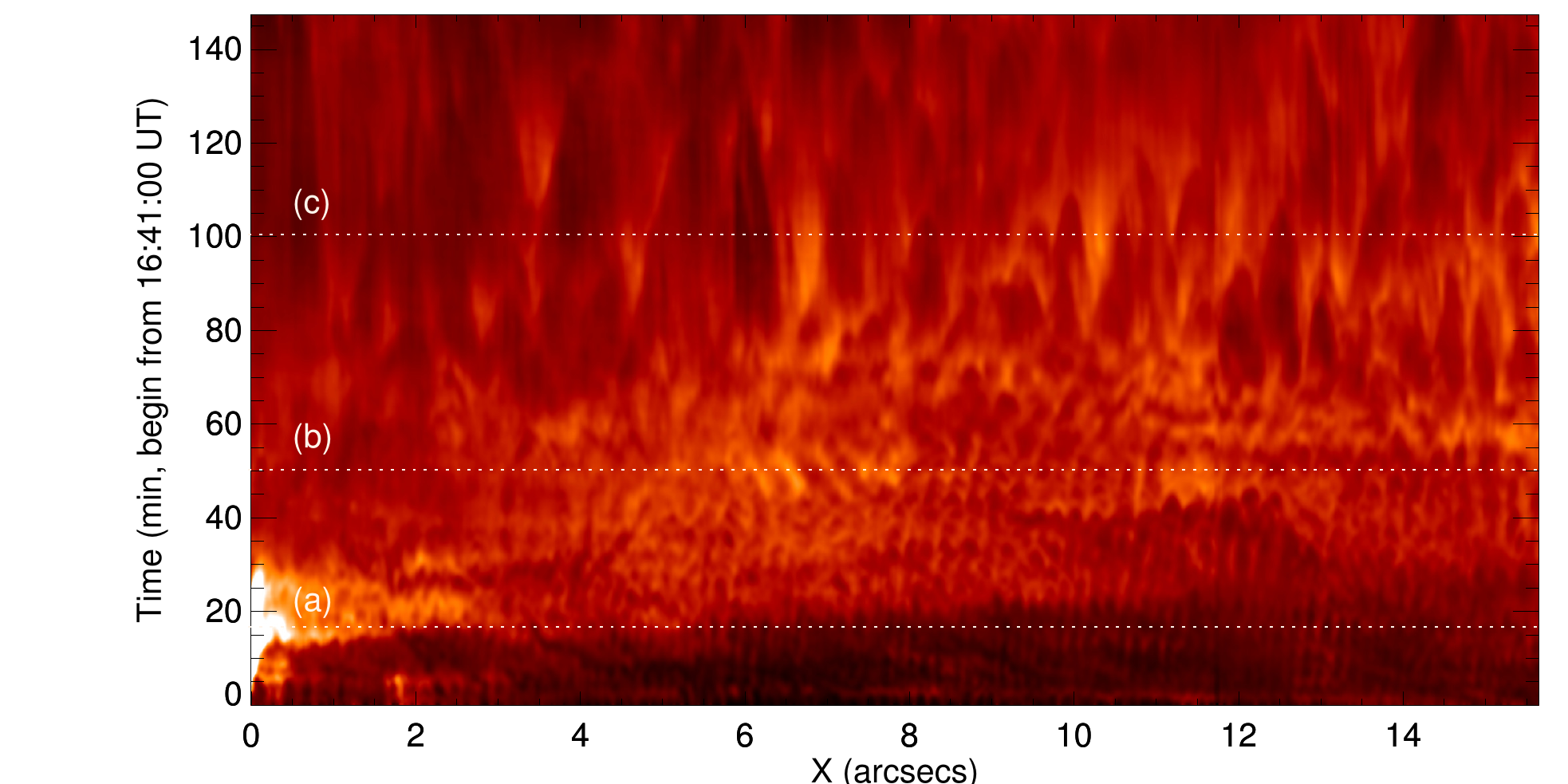}
            \caption{\normalsize Space time diagram of the H$\alpha$ line center image for DF marked as '1' in Figure 1. The three dotted lines indicates the umbral (a), penumbral (b) and superpenumbral (b) regions for which power spectrum is shown in Figure 7. }
            \label{Fig2}
            \end{figure*}
     \label{sect:des}
         
     A close look into the movie of H$\alpha$ line core images gives an insight into  fibrils located in the close proximity and in the vicinity of sunspot near the image center, a few of which were colligated with the superpenumbra. We investigate upon 40 such DFs for this study, wherein most AR DFs from the line core of H$\alpha$ images ensues almost a perfect parabolic paths. We opt for chosing the direction of a DF, or a bunch of DFs, manually using CRisp SPectral EXplorer, \citet[CRISPEX,][]{2012ApJ...750...22V} which helps in analyzing the multidimensional data cubes and an additional software called TANAT (Timeslice ANAlysis tools) for anlayzing and also for various measurements of data. A comprehensive use of this software not only made the seeking of sequence of images easier but also helped in tracking the events that were very clear while ascending and descending along their individual paths in 584 jet like features in H$\alpha$. The CRISPEX is a widget based versatile IDL tool for visual inspection and analysis of high resolution data. CRISPEX also provides  space-time diagram for all the linear as well as curved paths. This information could be stored and could be retrieved later for further analysis. We have produced the  space-time diagram for all the 40 DFs detected. The  space-time diagram for one of the DF is shown in Figure 2 and its evolution is shown in Figure 3. The figure shows the receding of the already risen DF along the same path. We have detected almost 584 such trajectories and did a parabolic fitting on it. Figure 4 gives an example of the fitting. From the fit, we calculated the parameters like deceleration, maximum velocity, maximum height, duration and kinetic energy as well.  We assume the electron density to be same throughout the chromosphere.
     This electron density  (${n_e}$) multiplied with the mass of the electron  (${m_e}$) gives
     density ($\rho$ = ${n_e}$ ${m_e}$ ) which means that the kinetic energy per unit volume would be
     equal to the square of the maximum velocity  (KE = ρ${v_{max}^2}$/2). The kinetic
     energy per unit volume of these jet like features may transform further into
     heat and might play a significant role in the heating of corona.

                       \begin{figure}
                       \centering
                       \includegraphics[width=13.0cm, angle=0]{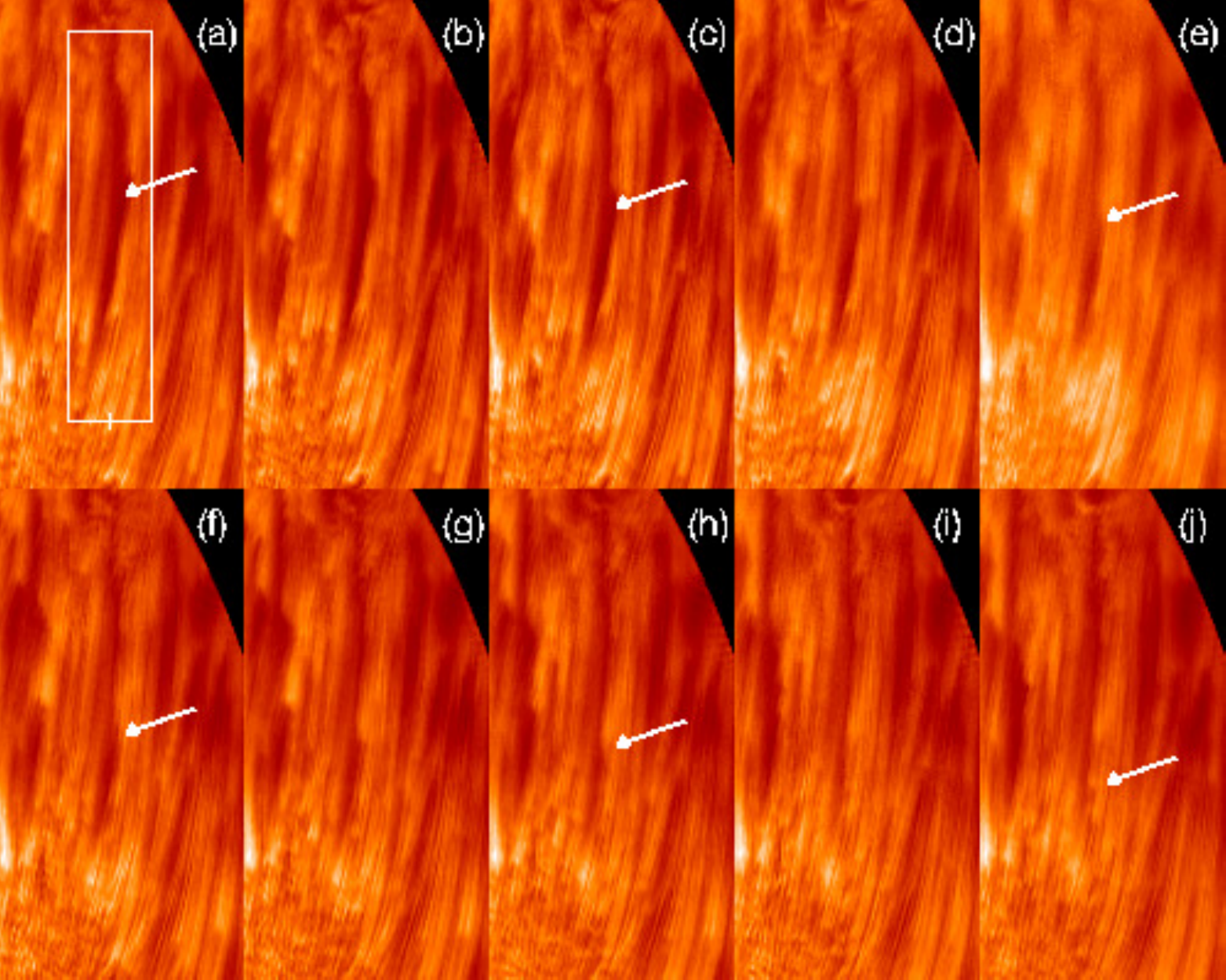}
                       \caption{\normalsize The snapshots of H$\alpha$ line core showing the temporal evolution of DFs (marked as '1' in Figure 1) as it recedes. [This event is available as an animation in the electronic edition of the Journal]}
                       \label{Fig3}
                       \end{figure}


                                \begin{figure*}
                                  \centering
                                  	\hskip 0.2cm
                                  \includegraphics[width=5.35cm,height=5.3cm, angle=0]{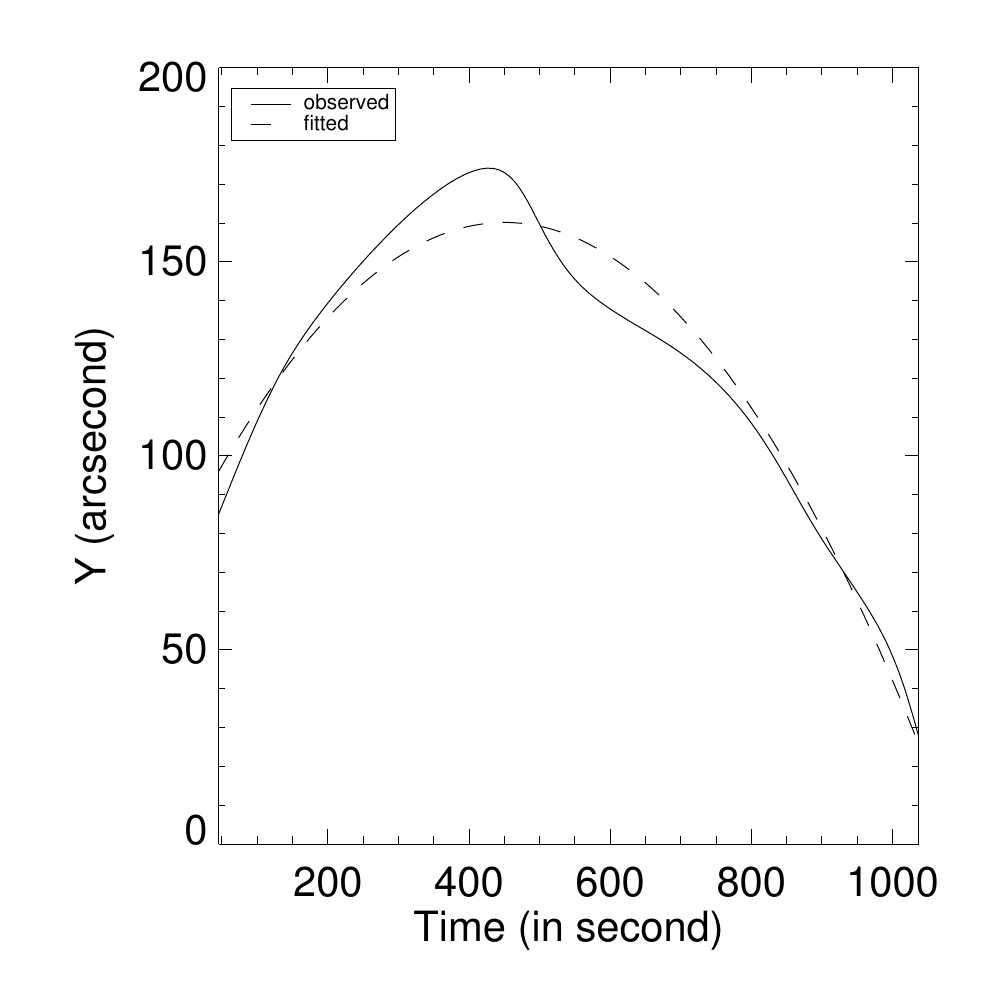} \includegraphics[width=5.35cm,height=5.3cm, angle=0]{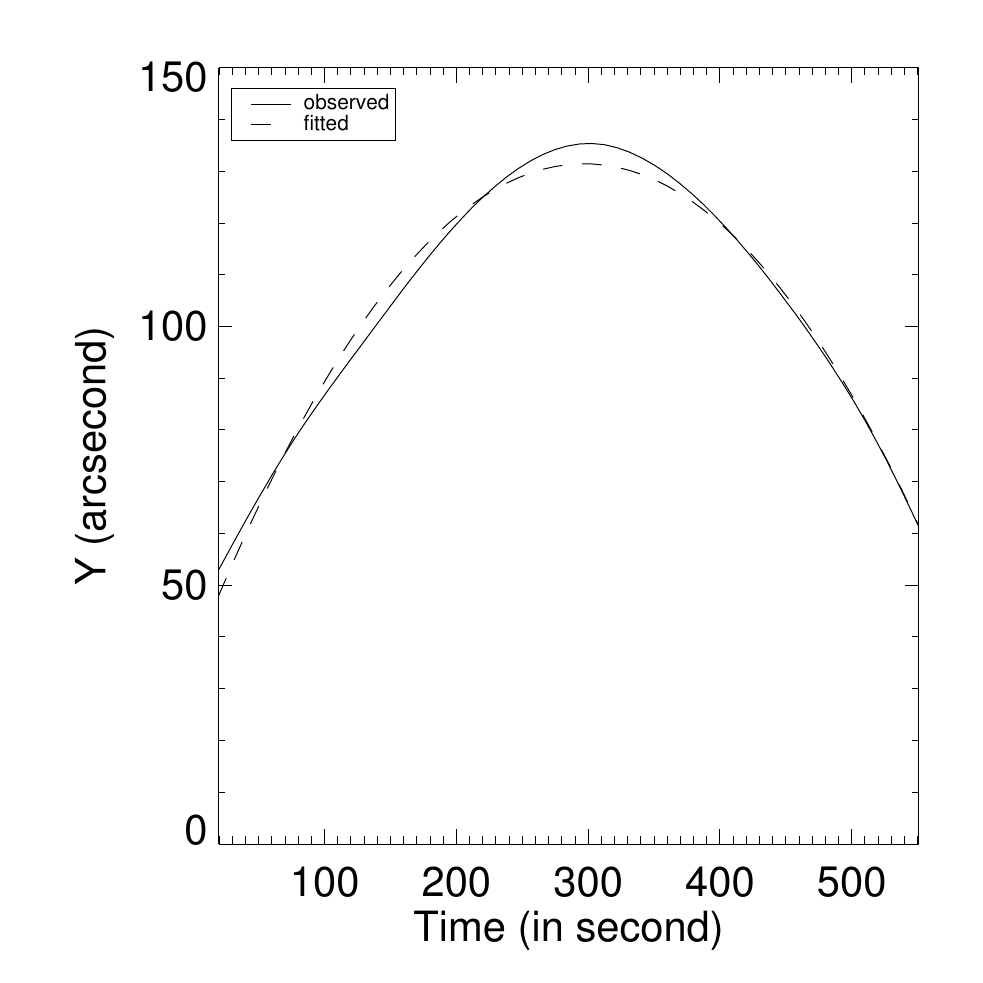} \\
                                 \includegraphics[width=5.2cm,height=4.8cm, angle=0]{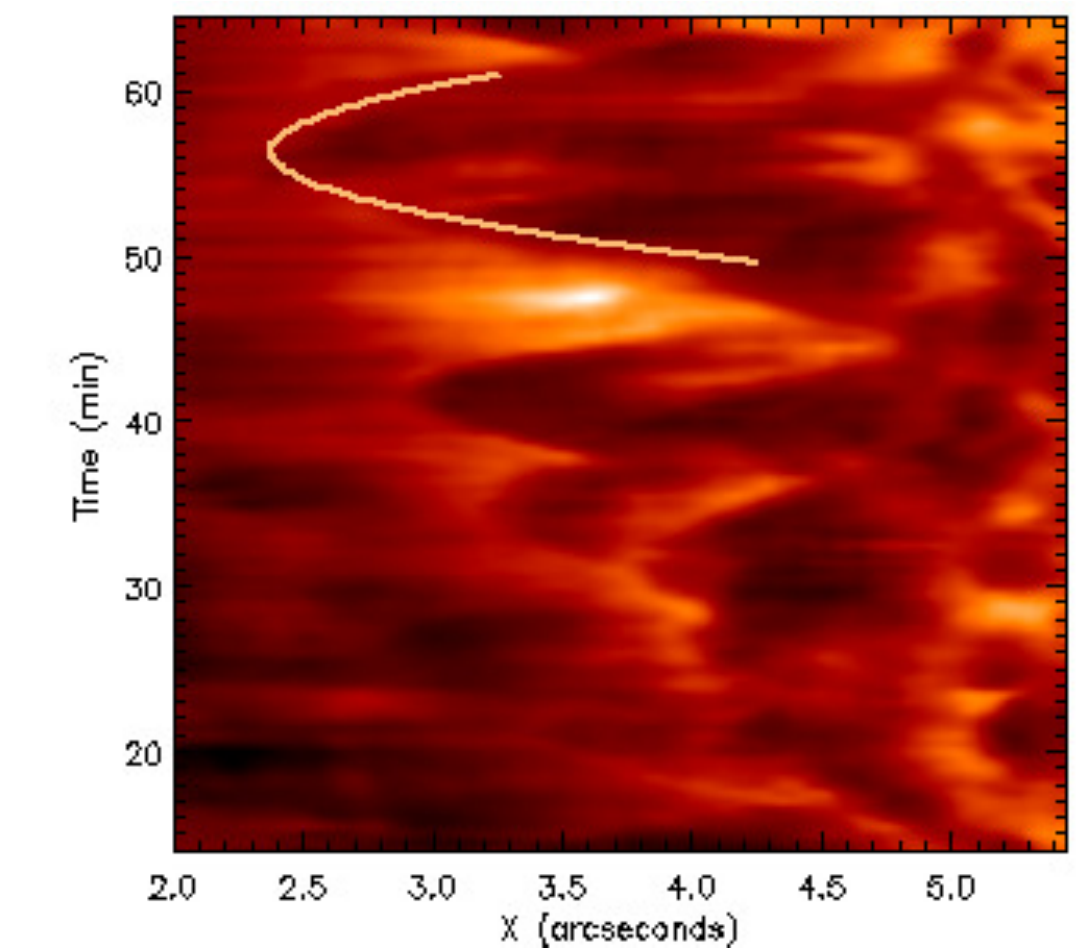}   \includegraphics[width=5.2cm,height=4.8cm, angle=0]{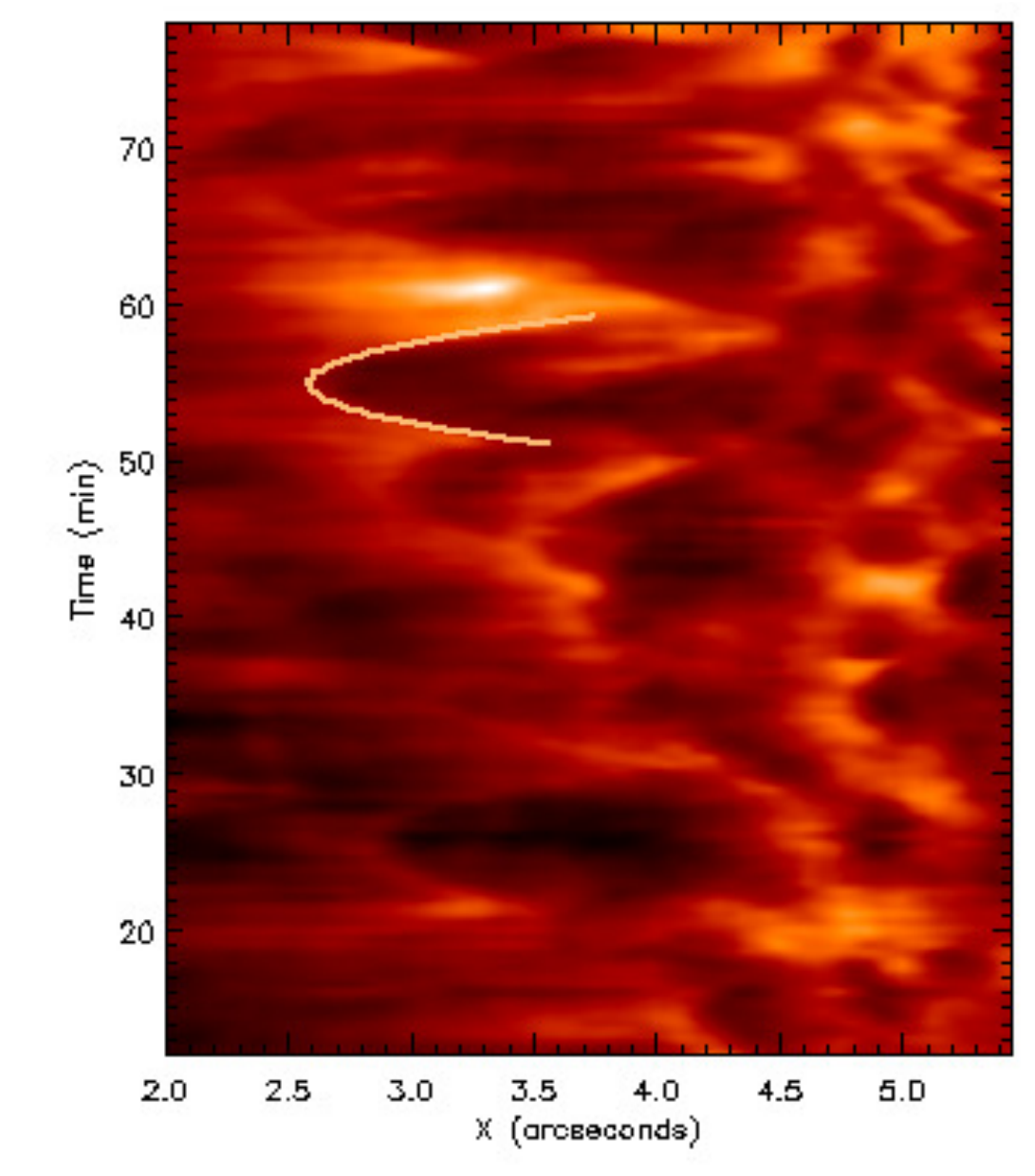}  
                                 
                                 \caption{\normalsize The extracted portion of x-t plot for sample DFs are shown in bottom panels. The DFs follows  almost perfect parabolic path. The yellow line indicates the bestfit used to derive the deceleration, maximum velocity, duration and the maximum height. The Parabolic fitting for corresponding trajectories are shown in top panels }
                                 \label{Fig4}
                                 \end{figure*}

                    \begin{figure*}
                    \centering
                   \includegraphics[width=9.0cm, angle=0]{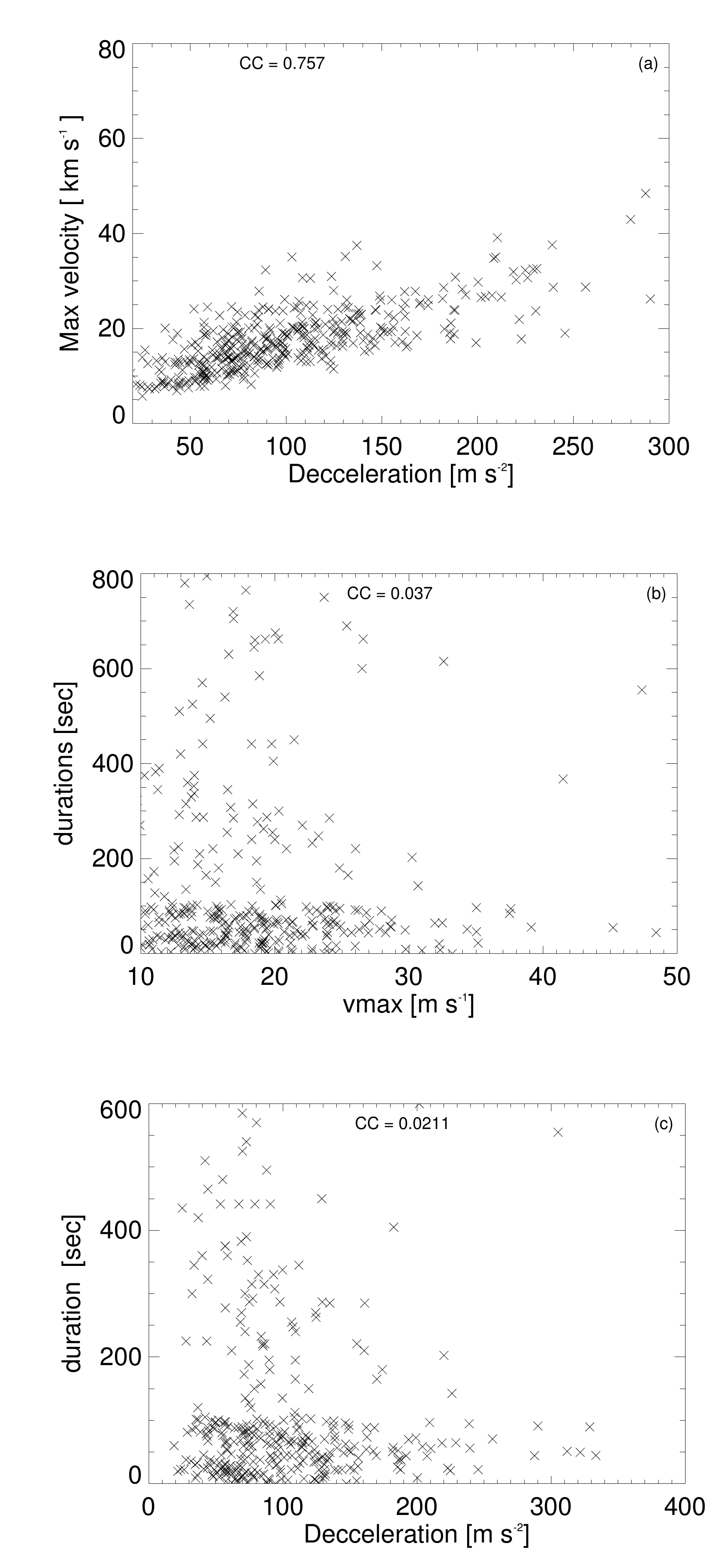}
                   \caption{\normalsize Scatter plots of deceleration vs. maximum velocity (a),  maximum velocity vs. duration (b), and decceleration vs. duration (C). A clear linear correlation exists between the deceleration and the maximum velocity.}
                  \label{Fig5}
                   \end{figure*}
                   \clearpage
                                                        
                 \begin{figure}
                 \centering
                 \includegraphics[width=15.0cm, angle=0]{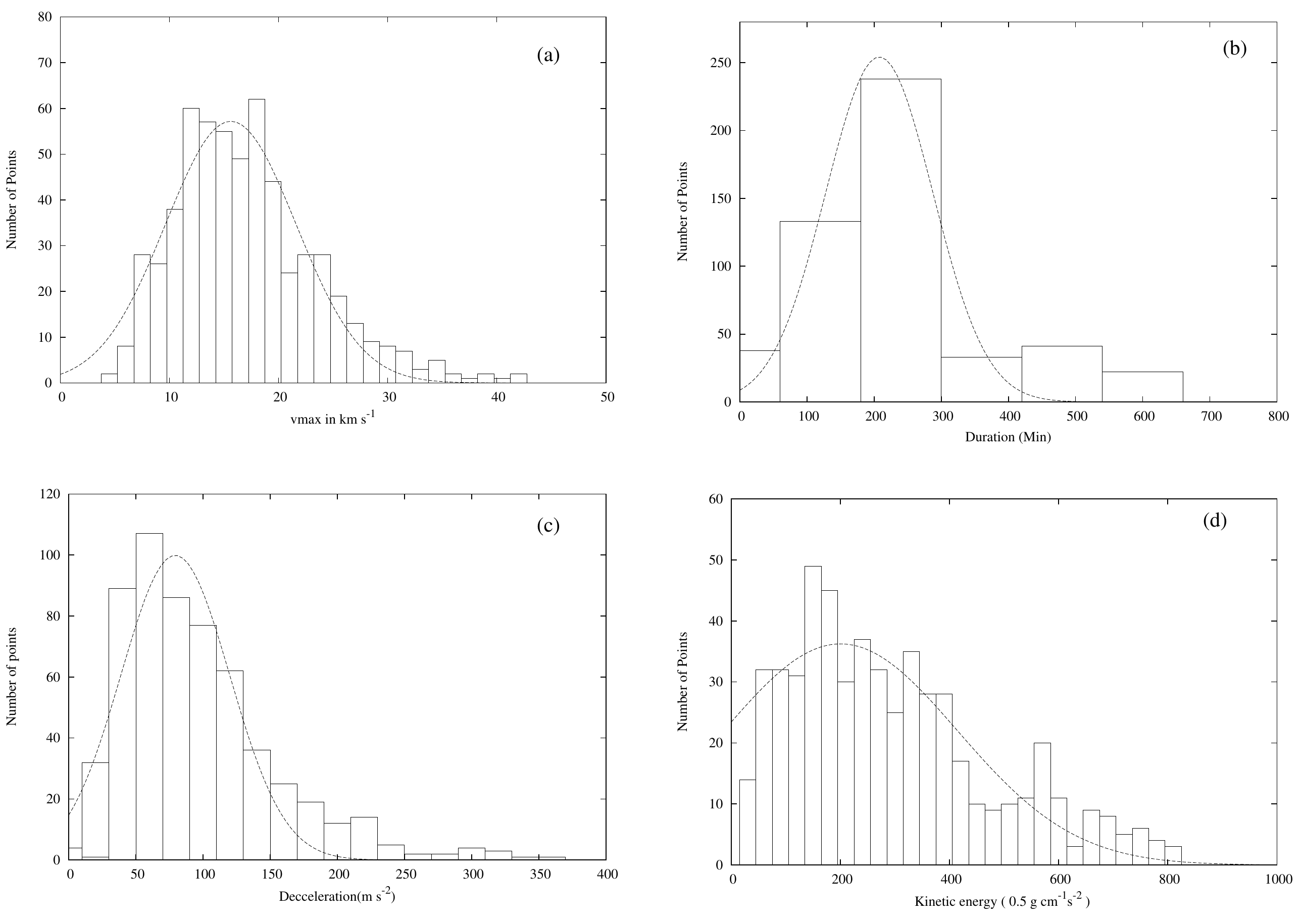}
                 \caption{\normalsize Distributions of the maximum velocity (panel a), duration (panel b), decceleration (panel c), Kinetic energy (panel d) respectively. The histograms are constructed from the observations of DFs.}
                 \label{Fig6}
                 \end{figure} 
                 \clearpage
                 
        \begin{figure*}
        \centering
       \includegraphics[width=14.0cm, angle=0]{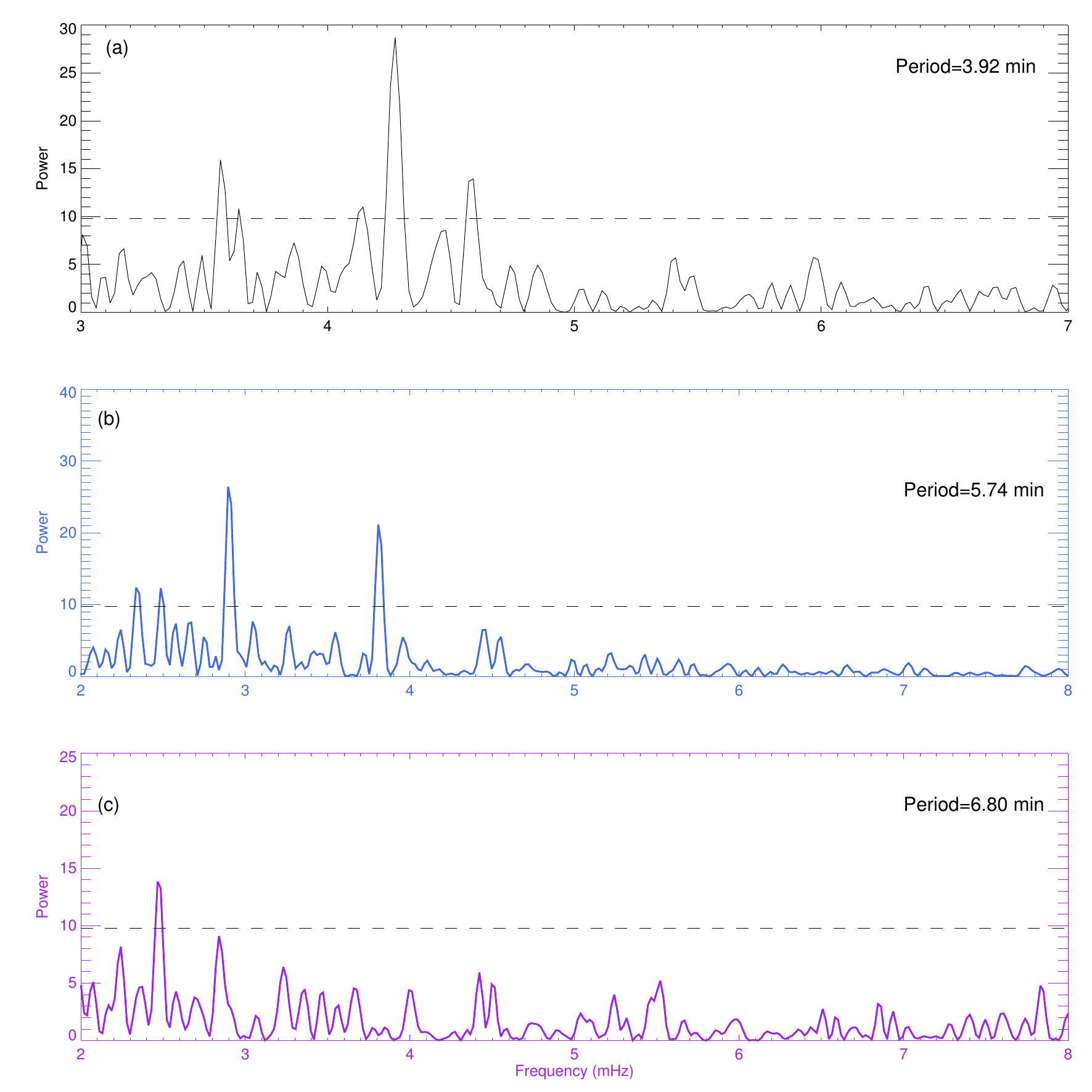}
       \caption{\normalsize The fourier power spectra at the positions marked in white dotted lines in Figure 2.}
       \label{Fig7}
       \end{figure*}
      
       We obtained the maximum velocity is between 10 km s$^{-1}$ and 30 km s$^{-1}$ and deceleration between 50 m s$^{-2}$ and 200 m s$^{-2}$. We noticed that  these correlations are quite similar to the measurements on dynamic fibrils done by \citet{2006ApJ...647L..73H}. 
        But we found that the correlation between the duration and the deceleration/maximum velocity is comparatively weaker, showing a high scatter and nonlinearity in almost all the DFs. 
                \begin{figure}
               \includegraphics[width=17.0cm, angle=0]{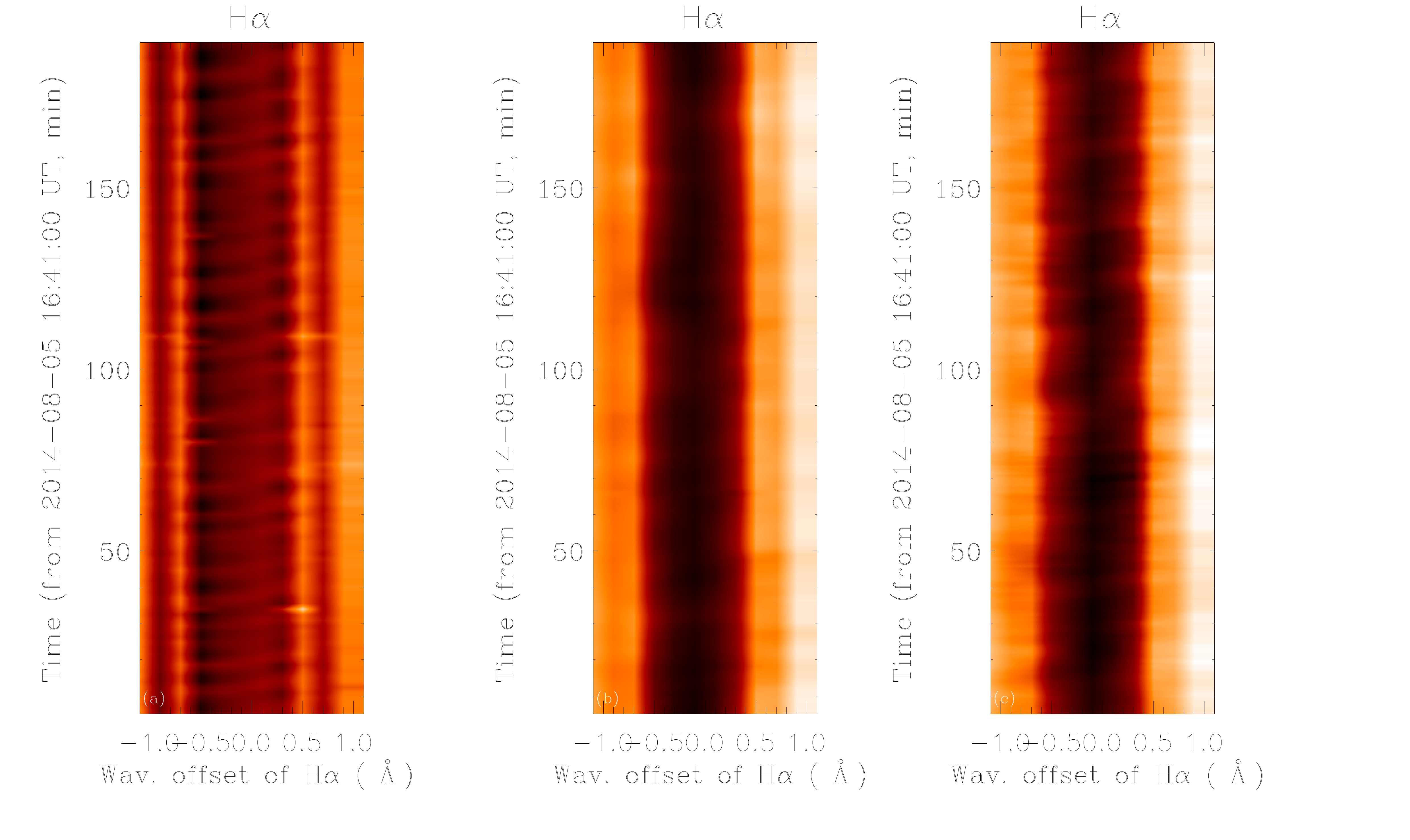}
              \caption{\normalsize Shock waves along the dynamic fibrils. The wavelength-time maps of the H$\alpha$ spectral line at different points along the DF as it propagates upward is shown in (a), (b) and (c)}
              \label{Fig8}
              \end{figure}  
      It could be seen in \citep{2007ApJ...655..624D} that the correlation between the duration and  the deceleration/ maximum velocity is not so clear or in other words weak. Similar results have been obtained by \citet{2008ApJ...673.1194L} in DFs with the reason being the effect of projection. It could be probable  that there is no effect of projection seen in duration whereas the effect is more prominent in velocity and deceleration. It is very clear from scatter plot in Figure 5 that the correlation between duration and deceleration/ maximum velocity is ambiguous with huge scatter and difference in slopes.
       By tracing all the jet like features we obtain the histograms of max velocity, duration, deceleration, and the kinetic energy shown in Figure 6. From the histograms, it is detected that the maximum velocity, duration and decelerations follows a Gaussian distribution. The maximum velocities are within 10 km s$^{-1}$ to 30 km s$^{-1}$ and the average is about 15 km s$^{-1}$, the duration ranges from 2 min to 20 min and the average is about 11 min, the deceleration ranges from 10 m s$^{-2}$ to 200 m s$^{-2}$ and the average is around 100 m s$^{-2}$. We obtain  the average maximum height to be around 0.025 Mm. We find the kinetic energy of all the 584 jet like features and estimated its average to be about 220 gm cm $^{-1}$s$^{-2}$ and found it to follow a partial Gaussian distribution. \\ 
       We take the fourier power spectra along the DF (originates from the umbral boundary and propagates higher up into the superpenumbral region) at three different points along the DF as shown in Figure 7. The power spectra in Figure 7(a) is at the umbral boundary, (b) is at penumbral region and (c) is at superpenumbral region. The running waves travels through the boundary of umbra and penumbra and eventually disappears at the penumbral boundary. In the power spectra we see oscillations in DF varying with time. To measure the oscillation period, we use the FFT (Fast Fourier Transform) function using IDL. As shown in Figure 7, the frequency at the umbral boundary is $\approx$ 4.3 mHz and based on the frequency of the power spectra, the oscillation periods at the same point is 235 s. The results show that the oscillations within umbra are a typical 3-min oscillation with period T = 235 s, and the
      oscillations of the same within penumbra are a  typical 5-min oscillation with period T = 345 s and T=408 s. The period of umbra oscillations is almost one half of that of the penumbral oscillations. The corresponding periods of umbral and penumbral oscillations measured by using FFT function are T$\_$umbra = 235 s and T$\_$penumbra = 408 s respectively. These oscillations propagates higher up and further develops into shocks as shown in Figure 8 suggesting  oscillations are upward propagating shock waves. Each panel in Figure 8 corresponds to three different positions along the marked DF in Figure  It is quite possible that the dynamic fibrils are usually driven in union with the sunspot oscillations \citep{2014ApJ...789..108C}. The Oscillations propagates higher up and develops into shock which propels the fibrils quite a several times. Thus, we speculate that the shocks driven DFs are related to the sunspot oscillations. To be more precise, there exists a form of propagation of shock wave fronts that starts from the sunspot center and reaches the bottom of the fibrils and propagates higher up along the DF \citep{2014ApJ...789..108C}. Hence, it is clear that the DFs are physically related to the sunspot oscillations and the form of shock fronts propagates from the sunspot center. Figure 8 shows the temporal-spectral variations (t-$\lambda$ plots) indicating  the oscillations at three different positions along the DF (connecting the sunspot to the fibril ) to infact be the upward propagating shock waves. \citet{2006ApJ...647L..73H} and \citet{2007ApJ...655..624D} simulated upward propagating  magnetoacoustic shock waves and created the dynamic fibrils observational behaviour. They found that the highly  dynamic chromospheric shock waves cause significant upward  and downward motion of the upper chromosphere. The transition of an upward propagating shock through the chromosphere produces a form of sawtooth or N-shape (as seen in Figure 8 ), which marks an impression of blueshift followed by a gradual drift towards the redshift and then a sudden appearance of  a blueshift \citep{2006ApJ...647L..73H,2009A&A...494..269V}. The N-shape pattern preponderate in all the three t-$\lambda$ plots constructed at different locations along the DF and hence confirms the idea of shock driven fibrils. The correlation coefficient between the maximum velocity and deceleration of a sample of 40 sunspot dynamic fibrils is 0.757. This proves that the oscillations in sunspots are the magnetoacoustic shock waves that propagates upward     (e.g.,\citealt{1986ApJ...301..992L,2006ApJ...640.1153C,2014ApJ...789..108C,2014ApJ...786..137T,2014ApJ...787...58Y}).\\ 
      
   \section{Discussion and conclusion}\label{sec:discussion}                                

   The above found correlations could probably be a significant signature for the jets that are periodic, are perhaps driven by the waves that normally propagates from photosphere into the chromosphere and hence steepen into shocks \citet{2013ApJ...776...56R}. The maximum velocities of these jet like features that we have found in our data are usually between 10$-$30 km s$^{-1}$ and is found to be similar in comparison to \citet{2006ApJ...647L..73H,2007ApJ...655..624D}. The chromospheric waves are usually created by the usual convective flows and also due to the oscillations found in photosphere and also in the convection zone. These disturbances further propagates into the chromosphere upwards and thereby shock and becomes a driving force for the plasma in chromospheric region and hence results in the formation of the DFs. 
    The periods that are longer than the period of local acoustic cut off is blocked by the chromosphere. This cut off period is completely dependent on the magnetic filed line inclination with respect to the vertical  \citep{1990LNP...367..211S,2004Natur.430..536D}. A single shock could possibly drive the DFs and is also the reason to understand the correlation between the deceleration and the maximum velocity. These DFs are the direct consequence of the upward propagating chromospheric oscillations/waves that are produced in the convection zone or photosphere as a result of global p-mode oscillations and also the convective flows. The waves propagates into the chromosphere and passes through upto the region where H$\alpha$ is formed, thus forming shocks thereby while its upward propagation. The fact on which we stress our interpretation is that the DFs follow a parabolic path with a good positive correlation between the properties like the velocity and the deceleration for all of the detected DFs. We found that velocity and deceleration are directly proportional with higher the velocity, the higher is the deceleration. The velocity that \citet{1995ApJ...450..411S} reports for their event of mottles is in the range of 10$-$30 km s$^{-1}$ and we found it to be rather similar to what we found in our BBSO/NST observations for our event of DFs. The same has been seen by \citet{2006ApJ...647L..73H} and is evident that the formation of DFs is due to the shocks in the chromosphere driven by the photospheric convective flows and oscillations as well. Hence, we conclude that the driving mechanism is same. \citet{2007ApJ...666.1277H}, \citet{2011ApJ...743..142H} simulation also explains that the usually long period waves propagates into the solar atmosphere along the magnetic field lines that are inclined. The most essential conclusion is with the explanation that the driving force for the DFs are the so called magnetoacoustic shocks that are induced by the p-mode oscillations and in addition leakage of convective flows into the chromosphere. Both the observations \citealt{1992ASIC..375..261L} and simulations \citealt{2010ApJ...722..888B} provides enough evidence for the waves that propagates upwards and also the shocks in the atmosphere of the sunspots. This phenomena appears as a dark feature in the H$\alpha$ images and are due to the density of magnetic flux.  \citet{2004Natur.430..536D} proposed that the similarity between the observations and some modeling done earlier on the DFs provides reason for the upward and downward movement of the upper chromosphere which is due to the shock waves in the chromosphere. \citet{2017ApJ...838....2Z} proves that the oscillations which appears to be like surge or light walls which lies above the light bridges are also caused by shocks . Our findings indicates that, in ARs,  most of these jet like features are caused by the shocks in chromosphere.

\begin{acknowledgements}

T.G.P would like to thank Patrick Antolin for his valuable help on CRISPEX.  This work is supported by the Grant 11427901, 11773038, 11373040, 11373044, 11273034, 11303048,11178005, AGS-0847126 and NSFC-1142830911427901 and also supported partly by  State Key Laboratory for Space Weather, Center for Space Science and Applied Research, Chinese Academy of Sciences. BBSO operation is supported by NJIT, US NSF AGS-1250818 and NASA NNX13AG14G. T.G.P thanks the financial support from CAS-TWAS Presidents PhD  fellowship - 2014. We thank referee for the review and for useful comments and suggestions.

\end{acknowledgements}


\begin{thebibliography}{40}
\providecommand\natexlab[1]{#1}
\providecommand\JournalTitle[1]{#1}

\bibitem[{Bard} \& {Carlsson}(2010)]{2010ApJ...722..888B}
{Bard}, S., \& {Carlsson}, M. 2010, \apj, 722, 888

\bibitem[{Beckers}(1968)]{1968SoPh....5..309B}
{Beckers}, J.~M. 1968, \solphys, 5, 309

\bibitem[{Bel} \& {Leroy}(1977)]{1977A&A....55..239B}
{Bel}, N., \& {Leroy}, B. 1977, \aap, 55, 239

\bibitem[{Cao} {et~al.}(2010)]{2010AN....331..636C}
{Cao}, W., {Gorceix}, N., {Coulter}, R., {et~al.} 2010, Astronomische
  Nachrichten, 331, 636

\bibitem[{Centeno} {et~al.}(2006)]{2006ApJ...640.1153C}
{Centeno}, R., {Collados}, M., \& {Trujillo Bueno}, J. 2006, \apj, 640, 1153

\bibitem[{Chae} {et~al.}(2014)]{2014ApJ...789..108C}
{Chae}, J., {Yang}, H., {Park}, H., {et~al.} 2014, \apj, 789, 108

\bibitem[{Christopoulou} {et~al.}(2001)]{2001SoPh..199...61C}
{Christopoulou}, E.~B., {Georgakilas}, A.~A., \& {Koutchmy}, S. 2001, \solphys,
  199, 61

\bibitem[{de Pontieu} {et~al.}(1999)]{1999SoPh..190..419D}
{de Pontieu}, B., {Berger}, T.~E., {Schrijver}, C.~J., \& {Title}, A.~M. 1999,
  \solphys, 190, 419

\bibitem[{de Pontieu} \& {Erd{\'e}lyi}(2006)]{2006RSPTA.364..383D}
{de Pontieu}, B., \& {Erd{\'e}lyi}, R. 2006, Philosophical Transactions of the
  Royal Society of London Series A, 364, 383

\bibitem[{De Pontieu} {et~al.}(2005)]{2005ApJ...624L..61D}
{De Pontieu}, B., {Erd{\'e}lyi}, R., \& {De Moortel}, I. 2005, \apjl, 624, L61

\bibitem[{De Pontieu} {et~al.}(2003{\natexlab{a}})]{2003ApJ...595L..63D}
{De Pontieu}, B., {Erd{\'e}lyi}, R., \& {de Wijn}, A.~G. 2003{\natexlab{a}},
  \apjl, 595, L63

\bibitem[{De Pontieu} {et~al.}(2004)]{2004Natur.430..536D}
{De Pontieu}, B., {Erd{\'e}lyi}, R., \& {James}, S.~P. 2004, \nat, 430, 536

\bibitem[{De Pontieu} {et~al.}(2007)]{2007ApJ...655..624D}
{De Pontieu}, B., {Hansteen}, V.~H., {Rouppe van der Voort}, L., {van Noort},
  M., \& {Carlsson}, M. 2007, \apj, 655, 624

\bibitem[{De Pontieu} {et~al.}(2003{\natexlab{b}})]{2003ApJ...590..502D}
{De Pontieu}, B., {Tarbell}, T., \& {Erd{\'e}lyi}, R. 2003{\natexlab{b}}, \apj,
  590, 502

\bibitem[{de Wijn} \& {De Pontieu}(2006)]{2006A&A...460..309D}
{de Wijn}, A.~G., \& {De Pontieu}, B. 2006, \aap, 460, 309

\bibitem[{Grossmann-Doerth} \& {Schmidt}(1992)]{1992A&A...264..236G}
{Grossmann-Doerth}, U., \& {Schmidt}, W. 1992, \aap, 264, 236

\bibitem[{Hale}(1908)]{1908ApJ....28..315H}
{Hale}, G.~E. 1908, \apj, 28, 315

\bibitem[{Hansteen} {et~al.}(2006)]{2006ApJ...647L..73H}
{Hansteen}, V.~H., {De Pontieu}, B., {Rouppe van der Voort}, L., {van Noort},
  M., \& {Carlsson}, M. 2006, \apjl, 647, L73

\bibitem[{Heggland} {et~al.}(2007)]{2007ApJ...666.1277H}
{Heggland}, L., {De Pontieu}, B., \& {Hansteen}, V.~H. 2007, \apj, 666, 1277

\bibitem[{Heggland} {et~al.}(2011)]{2011ApJ...743..142H}
{Heggland}, L., {Hansteen}, V.~H., {De Pontieu}, B., \& {Carlsson}, M. 2011,
  \apj, 743, 142

\bibitem[{Langangen} {et~al.}(2008{\natexlab{a}})]{2008ApJ...673.1194L}
{Langangen}, {\O}., {Carlsson}, M., {Rouppe van der Voort}, L., {Hansteen}, V.,
  \& {De Pontieu}, B. 2008{\natexlab{a}}, \apj, 673, 1194

\bibitem[{Langangen} {et~al.}(2008{\natexlab{b}})]{2008ApJ...673.1201L}
{Langangen}, {\O}., {Rouppe van der Voort}, L., \& {Lin}, Y.
  2008{\natexlab{b}}, \apj, 673, 1201

\bibitem[{Leighton} {et~al.}(1962)]{1962ApJ...135..474L}
{Leighton}, R.~B., {Noyes}, R.~W., \& {Simon}, G.~W. 1962, \apj, 135, 474

\bibitem[{Lites}(1986)]{1986ApJ...301..992L}
{Lites}, B.~W. 1986, \apj, 301, 992

\bibitem[{Lites}(1992)]{1992ASIC..375..261L}
{Lites}, B.~W. 1992, in NATO Advanced Science Institutes (ASI) Series C, Vol.
  375, NATO Advanced Science Institutes (ASI) Series C, ed. J.~H. {Thomas} \&
  N.~O. {Weiss}, 261

\bibitem[{Mart{\'{\i}}nez-Sykora} {et~al.}(2009)]{2009ApJ...701.1569M}
{Mart{\'{\i}}nez-Sykora}, J., {Hansteen}, V., {De Pontieu}, B., \& {Carlsson},
  M. 2009, \apj, 701, 1569

\bibitem[{Michalitsanos}(1973)]{1973SoPh...30...47M}
{Michalitsanos}, A.~G. 1973, \solphys, 30, 47

\bibitem[{Nishikawa} {et~al.}(1988)]{1988JGR....93.5929N}
{Nishikawa}, K.-I., {Frank}, L.~A., \& {Huang}, C.~Y. 1988, \jgr, 93, 5929

\bibitem[{Rouppe van der Voort} \& {de la Cruz
  Rodr{\'{\i}}guez}(2013)]{2013ApJ...776...56R}
{Rouppe van der Voort}, L., \& {de la Cruz Rodr{\'{\i}}guez}, J. 2013, \apj,
  776, 56

\bibitem[{Rutten}(2012)]{2012RSPTA.370.3129R}
{Rutten}, R.~J. 2012, Philosophical Transactions of the Royal Society of London
  Series A, 370, 3129

\bibitem[{Stangalini} {et~al.}(2012)]{2012A&A...539L...4S}
{Stangalini}, M., {Giannattasio}, F., {Del Moro}, D., \& {Berrilli}, F. 2012,
  \aap, 539, L4

\bibitem[{Suematsu}(1990)]{1990LNP...367..211S}
{Suematsu}, Y. 1990, in Lecture Notes in Physics, Berlin Springer Verlag, Vol.
  367, Progress of Seismology of the Sun and Stars, ed. Y.~{Osaki} \&
  H.~{Shibahashi}, 211

\bibitem[{Suematsu} {et~al.}(1995)]{1995ApJ...450..411S}
{Suematsu}, Y., {Wang}, H., \& {Zirin}, H. 1995, \apj, 450, 411

\bibitem[{Tian} {et~al.}(2014)]{2014ApJ...786..137T}
{Tian}, H., {DeLuca}, E., {Reeves}, K.~K., {et~al.} 2014, \apj, 786, 137

\bibitem[{Tsiropoula} {et~al.}(1994)]{1994A&A...290..285T}
{Tsiropoula}, G., {Alissandrakis}, C.~E., \& {Schmieder}, B. 1994, \aap, 290,
  285

\bibitem[{Tziotziou} {et~al.}(2004)]{2004A&A...423.1133T}
{Tziotziou}, K., {Tsiropoula}, G., \& {Mein}, P. 2004, \aap, 423, 1133

\bibitem[{Vecchio} {et~al.}(2009)]{2009A&A...494..269V}
{Vecchio}, A., {Cauzzi}, G., \& {Reardon}, K.~P. 2009, \aap, 494, 269


\bibitem[{Vissers} \& {Rouppe van der Voort}(2012)]{2012ApJ...750...22V}
{Vissers}, G., \& {Rouppe van der Voort}, L. 2012, \apj, 750, 22

\bibitem[{Yuan} {et~al.}(2014)]{2014A&A...561A..19Y}
{Yuan}, D., {Sych}, R., {Reznikova}, V.~E., \& {Nakariakov}, V.~M. 2014, \aap,
  561, A19

\bibitem[{Yurchyshyn} {et~al.}(2014)]{2014ApJ...787...58Y}
{Yurchyshyn}, V., {Abramenko}, V., {Kosovichev}, A., \& {Goode}, P. 2014, \apj,
  787, 58

\bibitem[{Zhang} {et~al.}(2017)]{2017ApJ...838....2Z}
{Zhang}, J., {Tian}, H., {He}, J., \& {Wang}, L. 2017, \apj, 838, 2

\end{thebibliography}

\end{document}